\documentclass[sn-mathphys,Numbered]{sn-jnl}

\usepackage{graphicx}%
\usepackage{multirow}%
\usepackage{amsmath,amssymb,amsfonts}%
\usepackage{amsthm}%
\usepackage{mathrsfs}%
\usepackage[title]{appendix}%
\usepackage{xcolor}%
\usepackage{textcomp}%
\usepackage{manyfoot}%
\usepackage{booktabs}%
\usepackage{algorithm}%
\usepackage{algorithmicx}%
\usepackage{algpseudocode}%
\usepackage{listings}%
\usepackage{orcidlink}

\usepackage{subcaption}

\theoremstyle{thmstyleone}%
\theoremstyle{thmstyletwo}%

\theoremstyle{thmstylethree}%

\begin{document}

\title[Article Title]{Assessment of few-hits machine learning classification algorithms for low-energy physics in liquid argon detectors}


\author*[1,2]{\fnm{Roberto} \sur{Moretti} \orcidlink{0000-0002-5201-5920}}\email{roberto.moretti@mib.infn.com}\equalcont{These authors contributed equally to this work.}

\author[3,4,5]{\fnm{Marco} \sur{Rossi}\orcidlink{0000-0002-7882-2798}}\equalcont{These authors contributed equally to this work.}

\author[1]{\fnm{Matteo} \sur{Biassoni} \orcidlink{0000-0002-9184-6217}}
\author[1,2]{\fnm{Andrea} \sur{Giachero} \orcidlink{0000-0003-0493-695X}}
\author[3]{\fnm{Michele} \sur{Grossi} \orcidlink{0000-0003-1718-1314}} 
\author[1,2]{\fnm{Daniele} \sur{Guffanti} \orcidlink{0000-0002-6182-5618}}
\author[1,2]{\fnm{Danilo} \sur{Labranca} \orcidlink{0000-0002-5351-0034}}
\author[1,2]{\fnm{Francesco} \sur{Terranova} \orcidlink{0000-0003-3044-7156}}
\author[3]{\fnm{Sofia} \sur{Vallecorsa} \orcidlink{0000-0002-7003-5765}}

\affil*[1]{\orgdiv{Dipartimento di Fisica ”G. Occhialini”}, \orgname{ Universit\`a di Milano - Bicocca}, \orgaddress{\city{Milan}, \postcode{I-20126}, \country{Italy}}}

\affil[2]{\orgname{Istituto Nazionale di Fisica Nucleare (INFN), Sezione di Milano-Bicocca}, \orgaddress{\city{Milan}, \postcode{I-20126}, \country{Italy}}}

\affil[3]{\orgname{European Organization for Nuclear Research (CERN)}, \orgaddress{ \city{Geneva}, \postcode{CH-1211}, \country{Switzerland}}}

\affil[4]{\orgdiv{Dipartimento di Fisica}, \orgname{Universit\`a degli Studi di Milano}, \orgaddress{\city{Milan}, \postcode{I-20133}, \country{Italy}}}

\affil[5]{\orgname{Istituto Nazionale di Fisica Nucleare (INFN) Sezione di Milano}, \orgaddress{\city{Milan}, \postcode{I-20133}, \country{Italy}}}

\abstract{The physics potential of massive liquid argon TPCs in the low-energy regime is still to be fully reaped because few-hits events encode information that can hardly be exploited by conventional classification algorithms. Machine learning (ML) techniques give their best in these types of classification problems. In this paper, we evaluate their performance against conventional (deterministic) algorithms. We demonstrate that both Convolutional Neural Networks (CNN) and Transformer-Encoder methods outperform deterministic algorithms in one of the most challenging classification problems of low-energy physics (single- versus double-beta events). We discuss the advantages and pitfalls of Transformer-Encoder methods versus CNN and employ these methods to optimize the detector parameters, with an emphasis on the DUNE Phase II detectors (``Module of Opportunity'').}




\maketitle

\section{Introduction}\label{sec:intro}

Liquid argon detectors play a prominent role in neutrino physics and - thanks to experiments like DUNE \cite{DUNE:2020lwj,DUNE:2020txw} and DarkSide \cite{DarkSide-20k:2017zyg,DarkSide:2014llq} - will be one of the technologies of choice for the next generation of accelerator neutrino experiments and direct search of dark matter. Such prominence is grounded on scalability. DUNE, in particular, has brought the liquid argon TPC technique (LArTPC) to an unprecedented scale after the development of cryostats that do not need to be evacuated to reach the required purity in TPCs \cite{Adamowski:2014daa}. In turn, this finding allowed the use of commercial membrane cryostats for the DUNE modules \cite{LBNE:2013lpy,Montanari:2015zwa,DUNE:2020cqd}. Similarly, DarkSide is commissioning high-throughput facilities for depleted underground argon extraction, purification, and distillation \cite{DarkSide-20k:2021nia,Alexander:2019uvv}.

In the last few years and, notably, in the course of the 2021 Snowmass process, several collaborations have formed to fill the gap between LArTPC for beam neutrinos and detectors for rare event searches \cite{DUNE:2022aul,Borkum:2023dsu,Back:2022maq,Avasthi:2022tjr,Parsa:2022mnj,caratelli2022lowenergy}. The first and second DUNE modules were engineered to achieve maximum performance for the observation of GeV-scale neutrinos but a wealth of DUNE physics resides in the observation of MeV events. Some of these channels are being actively pursued by DUNE because the event threshold $E$ is located at $E>10$ MeV. They are supernova neutrinos, sub-GeV atmospheric neutrinos, and the neutrinos originating from the sun from helium-proton fusion ("hep neutrinos") \cite{DUNE:2020zfm,DUNE:2020ypp}. Other channels are currently outside the DUNE scope but may be addressed by the DUNE Phase II modules (third and fourth modules). Recent studies conducted in the framework of the Module of Opportunity (MoO) R\&D effort \cite{moo_workshop_2022} indicate that a module with improved radiopurity, light collection efficiency, and granularity can access a rich physics portfolio: enhance the sensitivity to solar neutrinos produced by the $^{8}$B branch of the $pp$-chain, perform neutrinoless double-beta decay searches, boosted dark matter, and the direct detection of WIMP-like dark matter candidates \cite{Avasthi:2022tjr,Capozzi:2018dat,Mastbaum:2022rhw}. 

The possibility of making DUNE the most sensitive neutrinoless double-beta decay experiment in the world is still speculative because of the need for a large $^{136}$Xe mass to be dissolved in liquid argon with a few-percent concentration. Moreover, this search is hindered by the presence of radioactive isotopes of argon. Still, results on small prototypes that were doped at the level of 2\% are very encouraging \cite{CAMPESTRINI2014139}. Further, the ProtoDUNE-SP detector at CERN (700 tons of high-purity liquid argon) was doped in 2021 with $\simeq 100$ ppm and recorded no instability or performance deterioration \cite{Gallice:2021ykz}. The DUNE second module will be doped with (non-enriched) xenon even if the concentration of the double-beta decay isotope  $^{136}$Xe and the detector radiopurity and granularity cannot address the $^{136}\mathrm{Xe} \rightarrow ^{136}\mathrm{Ba}^* \ 2 e^-$ channel (Q-value: $\sim 2.458$ MeV) in a competitive manner. Several collaborations are addressing these challenges, which require high-throughput facilities for the extraction of underground argon depleted in both $^{39}$Ar and $^{42}$Ar \cite{Back:2022maq} or the development of dedicated systems for the distillation of atmospheric argon to remove $^{42}$Ar \cite{ar42distillation}. 

At the time of writing, the main obstacle toward large-mass LArTPCs in the few-MeV energy range is scalability. Charge readout in LArTPCs is carried out by wires or pixels, with pixels having been demonstrated to outperform the more traditional wire-based readout \cite{Adams_2020,Q-Pix}. Given the density of liquid argon, a few-MeV event will range out in $\sim 1$ cm and will produce just a few hits in the LArTPC. Bringing this number to a level comparable with supernova neutrinos would require a miniaturization of the charge readout system down to the diffusion scale ($\sim1$ mm) that is either impossible with wires or too expensive with pixels. In addition, the large increase in the number of readout channels impacts front-end electronics, data rate, and trigger complexity, and makes such a miniaturization approach cost-ineffective.

Machine learning (ML)-assisted algorithms have been proven to be the most effective choice to extract information in low-gra\-nu\-la\-ri\-ty detectors. Machine learning techniques were successfully applied to GeV neutrinos in LArTPC \cite{DUNE:2020gpm,SBNDGeV,sparseCNN,MicroBooNENN} and at lower energy in small-size devices \cite{Buuck:2022duk,ArgoNeuT}. A wealth of novel techniques for supervised and unsupervised machine learning already found applications in particle physics \cite{whitepaper_machinelearning} and they are particularly well suited to extract information when it is stored in variables that are non-trivially correlated, often surpassing the performances of non-ML approaches tailored for the task at hand. Few-hits low-energy events are thus an ideal target for these methods, which can compensate for the lack of detector granularity and relieve the burden of designing and operating a large number of channels in a cryogenic underground detector. 

This paper addresses such a challenge by identifying a class of machine learning approaches that are suited for low-energy LArTPC. The physics benchmark we employed in our study is the separation capability of one versus two electrons emerging from single and double-beta decay. We chose this benchmark because it represents an important but challenging handle to suppress background events from radioactive argon isotopes in any low-energy physics channel of LArTPCs. $\beta$ versus $\beta \beta$ separation is also useful to suppress $^{42}$K beta decay signals originating from $^{42}$Ar in the region-of-interest (ROI)  for the neutrinoless double-beta decay of $^{136}\mathrm{Xe}$. After identifying the optimal machine learning approaches, we show that these techniques relieve the requirement to invest in pixel miniaturization thus reducing the cost and complexity of the next-generation LArTPCs and of MoO. 

The main feature of LArTPCs of relevance for this study and the benchmark channels are introduced in Sec.\ref{sec:physics}. Sec.\ref{sec:machine_learning} presents the feature extraction and classification techniques for few-hits LArTPC events and the rationale of the different methods considered in this study. The performance of the methods against the benchmark channel ($\beta$-$\beta\beta$ separation) is discussed in Sec.\ref{sec:results}. In this section, we also discuss the impact on pixel miniaturization by comparing the overall effectiveness of deep learning algorithms with respect to a system that does not employ ML-assisted $\beta$-$\beta\beta$ separation techniques. We draw our conclusions in Sec.\ref{sec:conclusions}.

\section{Few-MeV events in LArTPC}
\label{sec:physics}

A liquid argon TPC is a cryogenic device that provides the full reconstruction of neutrino interactions in a broad energy range. The detectors that have been developed so far at large scale ($>100$ tons) are mostly based on wire anodes. Here, the electrons produced by ionization losses in LAr drift toward the anode driven by a constant electric field of $\sim$ 500 V/cm. The drift velocity at such a field amounts to 1.6 mm/$\mu$s and the electrons travel for several meters. In this paper, we will mostly consider the electric field configuration of the first DUNE module (FD1-HD), where electrons drift horizontally ("Horizontal Drift" - HD) between the cathode and the anode for a maximum distance of 3.5 m. 
FD1-HD is a 65.8 m $\times$ 17.8 m   $\times$ 18.9 m LArTPC segmented in four drift volumes for a total (fiducial) mass of 17 (10) kton. 
Electron recombination due to electronegative impurities has been addressed by decade-long R\&D. The purity of the argon achieved in a LArTPC is usually expressed in terms of the electron lifetime $\tau$. The lifetime is derived from the number of electrons that crosses a drift length $x=v_d t$ and it is given by:

\begin{equation}
    N(t)= N_0 \exp{-\frac{t}{\tau}} = N_0 \exp{-\frac{x}{v_d \tau}}
\end{equation}

where $N_0$ is the number of electrons (and ions) produced by a charged particle in a given volume of the detector, $x$ is the distance from the anode, $v_d$ is the drift velocity and $t$ is the drift time, i.e. the time the electron travels in LAr before reaching the anode. The electron lifetime thus characterizes the electron survival probability against electronegative impurities. In 2014, ICARUS observed a record lifetime of $>15 $ ms, corresponding to 20 parts per trillion (ppt) of O$_2$-equivalent contamination using a vacuum-tight cryostat \cite{Antonello:2014eha}. More recently, the DUNE Collaboration reached an even longer lifetime using a membrane cryostat. Data collected by the FD1-HD demonstrator (ProtoDUNE-SP) indicate a lifetime $>30$ ms over a maximum drift length of 3.5 m \cite{DUNE:2020cqd}. Such a bold result boosted the DUNE "Vertical Drift" concept that will be employed for the second DUNE module (FD2-VD). FD2-VD is based on a 10 kton LAr volume whose maximum drift length corresponds to 6 m. In the following, we will test our classification algorithms in a drift volume equivalent to one of the drift volumes of FD1-HD, properly accounting for electron losses due to residual impurities. Thanks to the outstanding purity reached by ProtoDUNE-SP, we anticipate that results hold for a drift length comparable with the maximum drift length of FD2-VD, too. 

The readout of ionization electrons at the cathode is generally performed by a set of wires that reconstruct the electron position in the anode plane, that is the plane perpendicular to the drift direction. In FD1-HD, charge reconstruction is performed by a set of 6 m $\times$ 2.3 m Anode Plane Assemblies (APAs). Each APA comprises four wire planes and the spatial resolution is dominated by the spacing of the wires inside the plane plus charge diffusion in LAr. FD1-HD employs 152 $\mu$m diameter copper-beryllium wires and the wire spacing on each layer is about $4.7$ mm, corresponding to a spatial resolution of $\sim 4.7 \ \mathrm{mm} /\sqrt{12}=1.36$ mm. FD2-VD will replace the APA wires with a pair of perforated PCBs, etched with readout strips. The collection strip corresponds to the strip where electrons are stopped and collected and it has a width of 5.1 mm. Induction strips crossed by the electrons before collection have a width of 7.65 mm. As a consequence, the space resolution of FD2-VD is comparable with FD1-HD. A novel readout based on pixels (pixel width: 4 mm) is employed in the DUNE near detector (NDLAr \cite{DUNE:2021tad}) and is being considered for the third and fourth DUNE far detector modules \cite{Parsa:2022mnj}. Unlike early LArTPCs, all DUNE modules will be operated employing front-end electronics operated at liquid argon temperature (87 K) because cold electronics boards located next to the readout element (wire, strip, or pixel) offer unprecedented noise immunity. The cold electronics of FD1-HD operates with noise well below 800$e^-$ per channel \cite{Adams:2020tqx}, corresponding to an energy threshold of $<50 $ keV. The front-end electronics for the pixelated system are under development and performance is expected to be comparable to or better than FD1-HD.

During the development and assessment of the ML-assisted event classifiers presented in this paper, the performance was estimated as a function of the spatial resolution and energy threshold within the range attainable by the technologies mentioned above.

Low-energy (1-10 MeV) events in liquid argon are recorded as a set of hits associated with an energy deposit per hit. The number of hits depends on the granularity of the LArTPC and, in particular, the size of the readout element (pixel or 2D hits reconstructed by the signals on wires). For a candidate neutrinoless double-beta decay of $^{136}$Xe (Q-value: $2.458$ MeV) it never exceeds 20 hits even in the most aggressive scenario (pixel size: 1 mm). Solar neutrinos offer a richer topology because the charged current interaction on $^{40}$Ar is accompanied by a de-excitation photon from the $\nu_e \ ^{40}\mathrm{Ar} \rightarrow ^{40}\mathrm{K}^* \ e^-$ reaction, while electron-neutrino scattering creates a single electron-like track only.

The ML-assisted identification techniques discussed below have been ranked against the most critical benchmark at the MeV scale: the identification of single versus double electrons in a given region-of-interest (ROI) when no other energy deposits are identified as a detached track (or hit) beyond the candidate electron track. The width of the ROI is determined by the energy resolution of the LArTPC, which exploits the total energy deposited estimated from the total collected charge and scintillation light. As a consequence, this information is not used to test the electron hypothesis. Further, we focused on backgrounds where the pulse-shape of the scintillation light cannot be exploited to identify the nature of the observed particle. This is a powerful technique in LAr for $\alpha-\beta$ and $n/\gamma$ separation using the time profile of the scintillation light \cite{Boulay:2006mb}. Still, for $\beta$ particles the identification can only rely on the hit topology and the pattern of ionization (charge) losses per hit. This situation comprises the two most critical backgrounds at the MeV scale, both originating from the radioactive isotopes of natural argon: single beta decay of $^{42}$Ar and pile-up events from $^{39}$Ar.

The presence of radioactive isotopes in natural argon is considered the most serious drawback of LAr detectors in low-energy physics and, in particular, to search for rare events like WIMP interactions, the occurrence of neutrinoless double-beta decay, and electron-neutrino scattering. Moderate-size LArTPC can be fil\-led with underground argon, which is depleted from radioactive isotopes, but the use of underground argon represents a challenge to the scalability of LArTPC to masses comparable with DUNE \cite{Andringa:2023aax}. In particular, $^{39}$Ar has a quite high natural abundance and contributes to an intrinsic activity of natural argon of $1.01\pm 0.08$  Bq/kg \cite{BENETTI200783}. This beta emitter has a lifetime of 269 y. Its Q-value (0.56 MeV) is immaterial for the physics processes considered in this paper (1-10 MeV) except in the occurrence of pile-up ($\beta \beta$ events). The same consideration holds for $^{42}$Ar, which contributes with a modest activity ($6\times 10^{-5}$
Bq/kg) and a sub-MeV Q-value (lifetime: 32.9 y). Unfortunately, the daughter isotope of $^{42}$Ar ($^{42}$K) is a beta emitter in secular equilibrium with  $^{42}$Ar and has a Q-value of 3.525 MeV. It thus represents the leading background to search for neutrinoless double-beta decay in DUNE \cite{Mastbaum:2022rhw}. 

ML-assisted identification algorithms are thus requested to separate single $\beta$  from double $\beta$ events exploiting the hit information mentioned above within a given ROI. For the sake of concreteness, we defined the true hypothesis considering the search for neutrinoless double-beta decay in DUNE. The signal is, therefore, the occurrence of two electrons with an energy deposit within an ROI centred at the Q-value of $^{136}$Xe. The hypothesis is tested against a single electron, whose energy is located inside the ROI.  
The comparison is performed among classification algorithms based on machine learning and compared with deterministic algorithms as the {\it blob} method developed by NEXT to address the same physics channel.

The performance of the classification techniques discussed in this paper is studied using two samples of simulated events:
one representing the background ($\beta)$, consisting of single electrons with energy equal to the $^{136}$Xe Q-value,
and a second made of $^{136}$Xe neutrinoless double $\beta$ decay events ($\beta\beta$) generated with energy and angular correlations as in \cite{deacy4}. 
The primary particles (single or double electrons) are then propagated inside a LAr volume using the Geant4 software package \cite{Geant4:2003,Geant4:2006,Geant4:2016} and the ionization energy loss simulated at each step is used 
to compute the number of ionization electrons to be propagated to the anode. We account for electron diffusion and recombination in liquid argon as described in Sec.\ref{sec:results}.
For simplicity, we consider in this work a pixel-based readout for it can highlight the impact of the system's spatial resolution
on the classification performance more intuitively than a wire-based anode. 

The pixel size and the lower limit on their threshold energy significantly affect the quality of $\beta$ and $\beta\beta$ events spatial reconstruction in a LArTPC. We accounted for these experimental limitations by spatially downsampling the energy deposition profiles in order to match a specific pixel size, removing the ones that don't satisfy the energy threshold requirement.

\section{Classification models and feature extraction}
\label{sec:machine_learning}
In this study, we considered three methods for classifying $\beta$ and $\beta\beta$ decays by using three-dimensional tracking information extracted from a LArTPC. The first method does not employ machine learning techniques and relies on a physics-informed extraction of highly discriminating features based on \textit{blob} detection, i.e. energy depositions in correspondence with the electron (or positron) trajectory endpoints. A variation of this technique, which we will refer to as ``blob method'', has already been applied by the NEXT Collaboration \cite{NEXT} for the $\beta$-$\beta\beta$ separation in a high-pressure $^{136}$Xe gas TPC with satisfactory results for tracks of $\sim 15$ cm \cite{Mart_n_Albo_2016,JRenner}. The second and third methods are Deep Learning architectures called Convolutional Neural Network (CNN) \cite{CNN} and Trans\-for\-mer-En\-co\-der, a variant of the Transformer \cite{aiayn}. CNNs are well-known models vastly applied in Computer Vision, while Transformers excel in Na\-tu\-ral Lan\-guage Pro\-ces\-sing prob\-lems thanks to the mechanism of self-attention. The NEXT Collaboration also developed CNN architectures \cite{NEXT2021} surpassing the blob method for background rejection in the $\beta\beta$ analysis. In the present work, CNN and Transformer solve the task of $\beta$-$\beta\beta$ binary classification with different feature processing strategies: the CNN analyses hit positions and energies as a set of pictures, while the Transformer treats hits energies and coordinates as sequences of correlated items. Moreover, the CNN specialises in learning from local features (i.e. pixel structures in a small neighbourhood) while the Transformer, due to its structure, captures both long and short-range dependencies equally \cite{atnsurvey}.\\
Within this framework, the blob method has been implemented to set a performance benchmark in class separation with respect to the Deep Learning models, as well as to investigate its limits at different granularities of the LArTPC charge readout. In the following, we describe in more detail the characteristics of the blob method and the Neural Network architectures employed in our analysis.

\subsection{The \textit{blob} method}
\begin{figure}
    \centering
    \includegraphics[width = 0.7\textwidth]{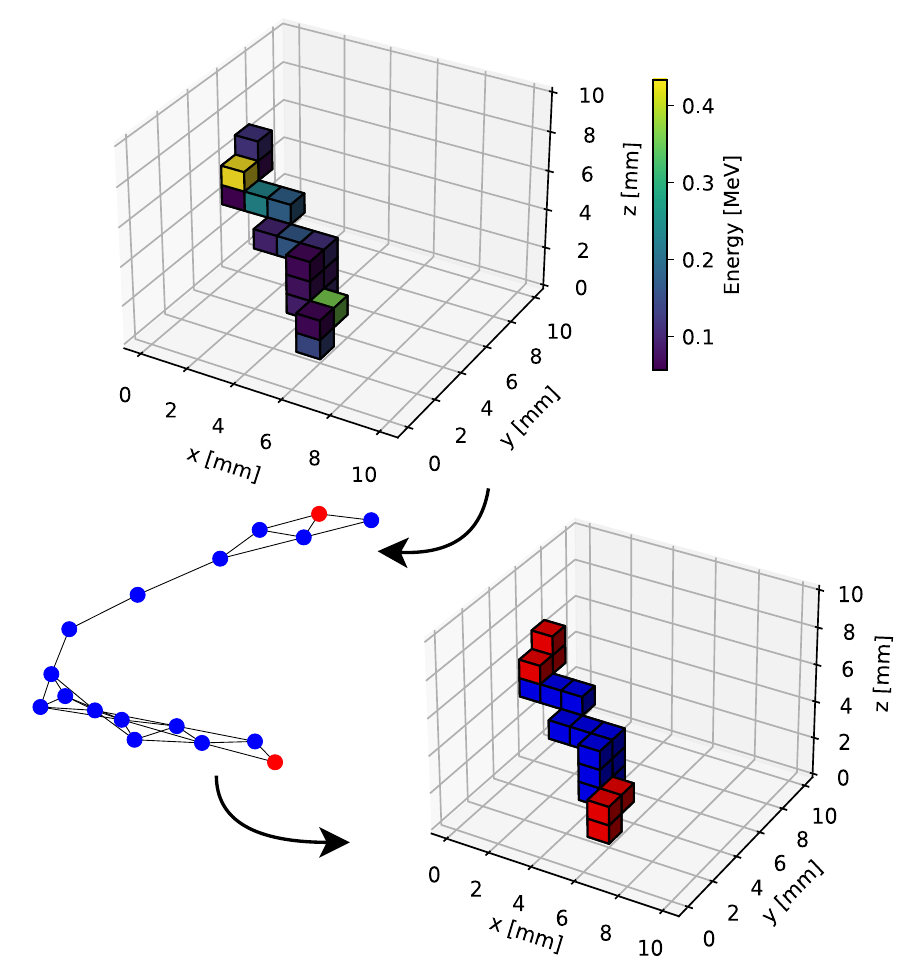}
    \caption{Workflow of the blob detection algorithm. The three-dimensional profile reconstructed at the LArTPC (top) is transposed into the corresponding graph. The red nodes represent the track endpoints, i.e. the candidate blob position (bottom left). We then integrate hit energies within a radius $r=2$ mm from the endpoints pair to determine $E_{b1}$ and $E_{b2}$ (bottom right).}
    \label{fig:1}
\end{figure}

\begin{figure}
    \centering
    \includegraphics[width = \textwidth]{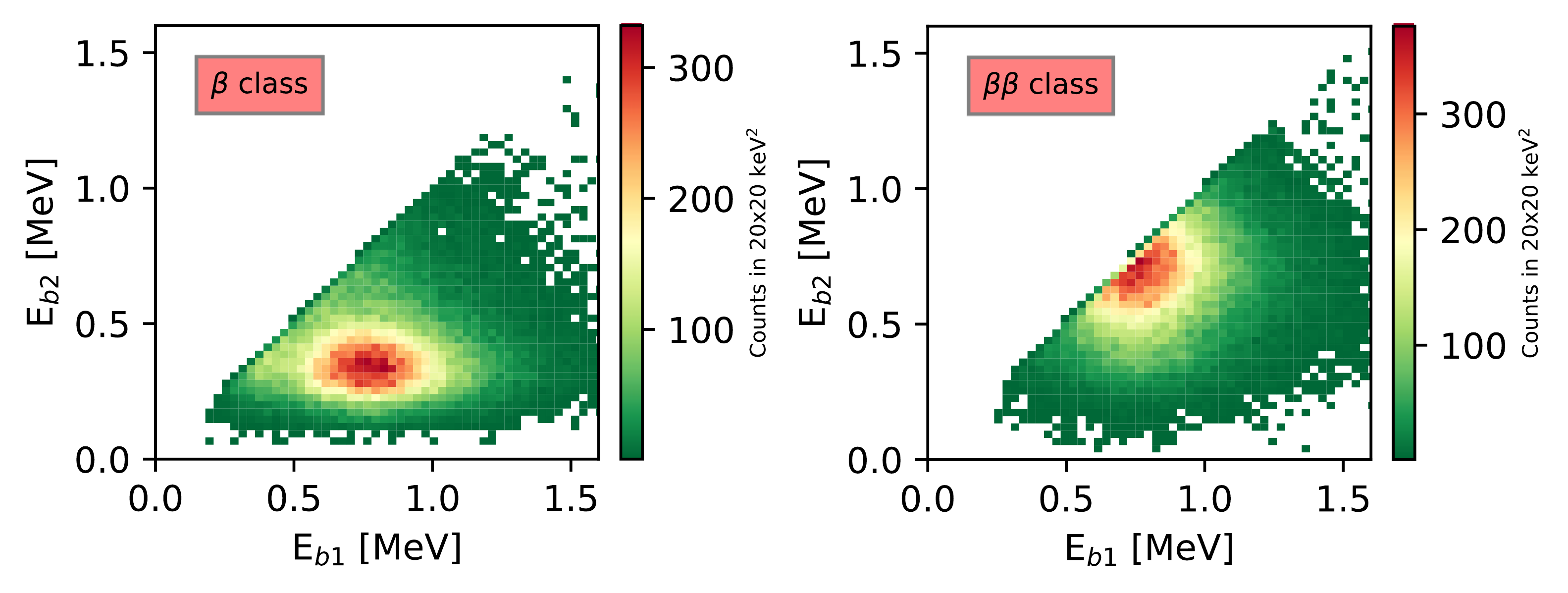}
    \caption{Blob candidate energy distributions for the $\beta$ class (top) and the $\beta\beta$ class (bottom). $E_{b1}$ and $E_{b2}$ are extracted from the three-dimensional LArTPC event reconstruction considering a pixel size of $1\times1\times1$ mm${}^3$ and a hit energy threshold of $50$ keV. As expected, the $\beta\beta$ distribution centroid appears closer to the bisector than the $\beta$'s one, allowing for class separation.}
    \label{fig:2}
\end{figure}

Due to the inverse-square velocity dependence of the average ionization energy loss per unit distance \cite{pdg}, $\sim 1$ MeV electrons release more of their energy when close to their trajectory endpoint, forming a blob. This implies that the double-beta ionization pattern will appear as a unique track in the LArTPC with two endpoint blobs, whereas single-betas only feature one endpoint. We detected blob candidates by finding an appropriate graph representation for each event and localizing the nodes corresponding to the blob position using a Breadth-First Search (BFS) \cite{Silvela2001BreadthfirstSA}. For a $n$-hits track, the algorithm works as follows:
\begin{enumerate}
    \item Assign every track hit to a graph node and connect every node pair corresponding to adjacent hits with edges of unitary weights. Two hits are considered adjacent if they share a surface, an edge or a vertex in the three-dimensional lattice.
    \item Perform a BFS search to find the shortest path length between each pair of nodes in the graph $i, \,j$ for $i,\,j=1,\,...\,n$. Let $D$ be the symmetric $n\times n$ matrix collecting the pairwise path length.
    \item Find the indices $i' j'$ such that $D_{i'j'} \geq D_{ij} \; \forall \; i=1,\,...\,n \; \land \; i<j<n$. 
    \item Associate the track endpoints (i.e. the candidate blobs centres) to the positions of hits $i'$ and $j'$.
    \item Sum the energies of all hits within a blob radius $r$ from each centre, as depicted in Fig. \ref{fig:1}, obtaining the new variables $E_{b1}$ and $E_{b2}$, where $b_1$ is the more energetic blob candidate: $E_{b1} > E_{b2}$.
\end{enumerate}
For $\beta\beta$ events, both candidates are expected to be true blobs, hence $E_{b1}\simeq E_{b2}$. On the contrary, for $\beta$ decays $E_{b2}$ should be significantly smaller than $E_{b1}$. These two variables allow for establishing a two-feature background rejection criterion. By leveraging the Neyman-Pearson lemma \cite{npl}, we employ the likelihood ratio as the test statistics:
\begin{equation}
    q = \frac{P(E_{b1}, E_{b2}\, | \, \beta\beta)}{P(E_{b1}, E_{b2} \, |\, \beta)}
\end{equation}
where $P(E_{b1}, E_{b2}\, | \, \beta\beta)$ and $P(E_{b1}, E_{b2} \, |\, \beta)$ are the probability distributions for an event with $E_{b1}, E_{b2}$ under the $\beta\beta$ and $\beta$ hypothesis, respectively. These probabilities are unknown \textit{a priori} and we performed a data-driven estimation by sampling the $E_{b1}$, $E_{b2}$ distributions in the training dataset. Fig. \ref{fig:2} shows an example of $E_{b1}$ and ${E_{b2}}$ feature distributions for the $\beta$ and $\beta\beta$ classes.

Despite its simplicity, the blob method has two drawbacks:
\begin{itemize}
    \item part of the track information is lost, i.e. only hits near the endpoints deliver feature information.
    \item the BFS fails in determining blob positions if the track reconstructed by the LArTPC presents many gaps, for example due to inefficiencies or trigger requirements, like setting a lower limit energy threshold for the hits, as described in Sec.\ref{sec:results}.
\end{itemize}

\noindent In addition, this method cannot be generalized to background reduction for other physics channels.

\subsection{Convolutional Neural Network}
The fundamental building block of CNNs is the convolutional layer, which is a set of back-propagation learnable filters, i.e. tensors which perform a convolution operation on a fixed-size input. Thanks to a stack of convolutional layers, a CNN is able to process hierarchical features that are significant for the learning process \cite{Bilal_2018}.
In addition to convolutional layers, CNNs typically include pooling layers, which downsample the feature maps to reduce the computational training cost of the network and limit overfitting. Fully connected layers are then added to the network's end to map the high-level features to the desired output.

\begin{figure*}
    \centering
    \includegraphics[width = 0.9\textwidth]{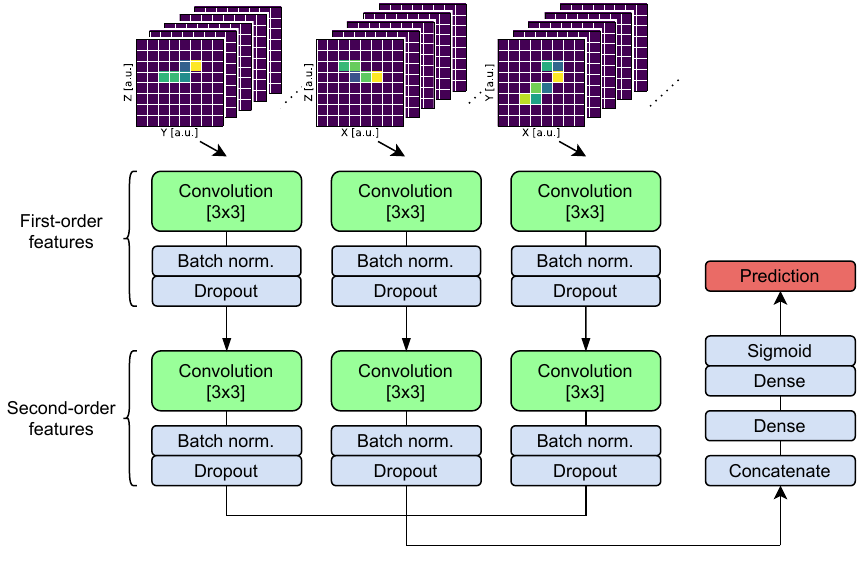}
    \caption{Convolutional Neural Network scheme. A batch of LArTPC events split into three planar views are fed to two independent stacks of convolutional, batch normalization \cite{bnorm} and dropout \cite{dropout} layers. The stack outputs merge into a single array, which passes through a fully connected layer. The output layer is a single neuron with a sigmoid activation function, which returns a $[0, \, 1]$ bounded network predictive score. Each convolutional step comprises $25$ filters with $[3\times3]$ dimensions, and all hidden layers are equipped with LeakyReLU activations \cite{leakyrelu}. Dropout layers were inserted to prevent overfitting of the model.}
    \label{fig:3}
\end{figure*}

In order to define a scalable CNN architecture for the task of track classification in LArTPC when higher readout granularity rapidly increases the input dimension, we embedded the hit energy content into a three-dimensional tensor according to their position in cartesian coordinates, setting all the other entries to zero. We then integrated along the orthogonal axis X, Y, Z to get the three planar views (YZ, XZ and XY planes, respectively). At the cost of a  dispensable loss of information, this allows the CNN to support a wide range of readout resolutions with fixed architecture and hyperparameters, with manageable computational costs and resources (only two-dimensional filters are needed), despite employing a large training dataset, containing about $2\times 10^5$ events. In this physics application, we were also able to remove the pooling layers, which compromise the performance for the low granularity configurations. Fig. \ref{fig:3} illustrates a schematic representation of the CNN architecture we designed for this study.

\subsection{Transformer and self-attention}
The attention mechanism is the core of Transformers \cite{aiayn}, which allows the model to adaptively focus on specific parts of the input sequence, i.e. hits that are supposed to carry more information than others for the classification task. In particular, we refer to the Scaled Dot-Product attention \cite{aiayn} shown in equation \ref{eq:3}.
Given an ensemble of input sequences, the self-attention mechanism computes a set of query (Q), key (K), and value (V) matrices for each input sequence via fully-connected layers. These matrices are then used to compute a weighted sum of the values, where the weights are determined by the similarity between the query and key matrices. More specifically, the attention weights for a given query matrix are computed as a softmax function \cite{softmax} over the dot products of the query and key matrices. The resulting weights are then used to compute a weighted sum of the value matrices, resulting in a context vector \cite{contextvector1,contextvector2} that captures the most relevant information for the query.

The self-attention mechanism is often used in a multi-head configuration, where multiple sets of query, key, and value matrices are computed in parallel. The resulting context vectors from each head are concatenated and passed through a linear layer, which combines the information from the different heads.

Transformers typically use an encoder-decoder architecture, which consists of an encoder network that processes the input sequence and a decoder network that generates an output sequence. The encoder and decoder both use stacked self-attention layers followed by feedforward layers and are connected by means of an additional attention layer that computes the context vector based on the encoded input sequence. 

\begin{figure*}
    \centering
    \includegraphics[width = 0.8\textwidth]{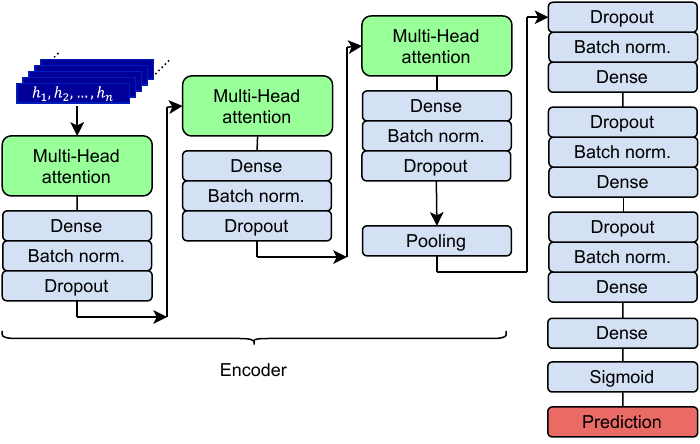}
    \caption{Transformer-Encoder scheme. LArTPC events consist of a collection of hits ($h_1\, h_2, ...\, h_n$), where $n$ can change for different events. Every hit comes with four variables (three space coordinates and the hit energy), and the input vector size is $4m$, where $m$ is the largest hit number for the events in the batch. The encoder consists of three stacks of multi-head attention layers and fully-connected layers followed by batch normalization and dropout. Each multi-head step comprises four parallel self-attention heads. The encoder output is mapped into the final prediction, i.e. a single-neuron layer with a sigmoid activation function. All hidden layers are equipped with LeakyReLU activations.} 
    \label{fig:4}
\end{figure*}

For binary classification tasks, the network's output simply consists of a single prediction value. For this reason, we employed a simplified architecture consisting only of the encoding part of the Transformer, with a stack of feed-forward layers mapping the encoded state to the output. In this architecture, only self-attention is needed. Self-attention is computed by:
\begin{equation}
    \text{Attention(Q, K, V)} = \text{softmax}\left(\frac{QK^T}{\sqrt{d_k}} \right) V .
    \label{eq:1}
\end{equation}
$Q=K=V$ are matrices $\in \mathbf{R}^{d_{\text{model}} \times d_{k}}$, where $d_{\text{model}}$ is the embedding size and $d_{k}$ the sequence length. The Transformer-Encoder we developed for this work is depicted in Fig. \ref{fig:4}.
The Transformer-Encoder takes as input the full three-dimensional information. LArTPC events are thus treated as weighted point clouds in which every point corresponds to a hit position in space and the weight is determined by the hit energy. The hit energy and position are fed to the network as a unique array per event with four entries per hit (XYZ coordinates and its energy E). Zero-padding is required for every training batch in order to preserve dimensions through a forward-backwards pass.
Compared to the traditional CNN, this Transformer implementation allows for more efficient memory management, with faster training at higher spatial resolutions. This is due to the fact that CNN needs to store progressively larger and sparser two-dimensional tensors and suffers from the increase in dimensionality. The Transformer-Encoder might emerge as a potential rival to tailored approaches such as sparse CNNs, which already overcome the memory issues of CNNs in the classification of GeV-scale events in LArTPCs \cite{sparseCNN}.

\section{Dataset and results}
\label{sec:results}
Particle ionization losses in liquid argon produce a number of charge carriers (electrons and ions) that is proportional to the deposited energy. The carriers drift across the medium thanks to the TPC electric field. LArTPCs enable three-dimensional tracking by recording the signal induced by ionization electrons as they approach the anode plane. The quality of the event reconstruction depends on the granularity of the readout system, i.e. the anode wire pitch or pixel size, and the sampling rate of the induced signal. The latter determines the space precision along the drift coordinate $Z$. Few-MeV tracking is also sensitive to the minimum detectable charge by the front-end electronics as discussed in Sec.\ref{sec:physics}. This limitation corresponds to an energy threshold  of several tens of keV per hit.

In this work, we considered the same signal sampling rate as DUNE ($2$ MHz \cite{DUNE:2020lwj}), which corresponds to $\sim1$ mm spatial resolution in the drift direction for an electric field of $E=500$ kV/cm. We then trained the classification models introduced in Sec.\ref{sec:machine_learning} by varying the pixel size $w$ at different energy thresholds $E_t$.

The dataset consists of $2\times 10^5$ events, equally split into the $\beta$ and the $\beta\beta$ classes. Event distances from the readout plane are uniformly distributed between $0$ m and the maximum drift length ($3.5$ m). Electron diffusion was taken into account by applying longitudinal and transversal dispersions according to:
\begin{equation}
    \sigma_L = \sqrt{2D_Lt}
    \label{eq:3}
\end{equation}
\begin{equation}
    \sigma_T = \sqrt{2D_Tt}
    \label{eq:4}
\end{equation}
where $D_L$ and $D_T$ are longitudinal and transversal diffusion coefficients estimated for liquid argon from empirical models \cite{Li_2016} and $t$ is the drift time. We estimated a maximum standard deviation of $\sigma_L \approx 1.7$ mm and $\sigma_T \approx 2.3$ mm, corresponding to an event occurring at $3.5$ m distance from the readout (maximum drift time). As noted in Sec.\ref{sec:physics}, we accounted for electron recombination by assuming an effective electron lifetime of $\tau = 30$ ms.
Note that recombination plays a marginal role in our study since the maximum drift time ($\sim 2.2$ ms) is much smaller than the electron lifetime. This implies that the majority of electrons will reach the anode before undergoing recombination.

For each training at a different pixel size and energy cutoff, we downsampled the MC simulation track information by integrating the energy depositions into hits of dimension $w\times w \times 1$ mm removing hits below the energy threshold. We trained each model by randomly partitioning the dataset into $140000$ events for training ($70\%$), $30000$ for validation ($15\%$) and $30000$ for testing ($15\%$). To establish the model performance, we chose the accuracy metric as the fraction of events correctly classified by the model. For a balanced dataset, i.e. with the same number of samples for each label, the accuracy value ranges between the rates of true $\beta\beta$ event acceptance and the true $\beta$ event rejection.
The CNN and the Trasformer-Encoder training was carried out with the Adam optimizer \cite{kingma2017adam}, a variant of the Stochastic Gradient Descent (SGD), with an initial learning rate of $10^{-3}$. The learning rate halves every $20$ consecutive stall epochs, i.e. epochs in which the validation accuracy does not increase with respect to the previous one. $50$ epochs for the Transformer-Encoder and $40$ for the CNN ensure a stable convergence of the learning curves. 

\begin{figure}
    \centering
    \includegraphics[width=\textwidth]{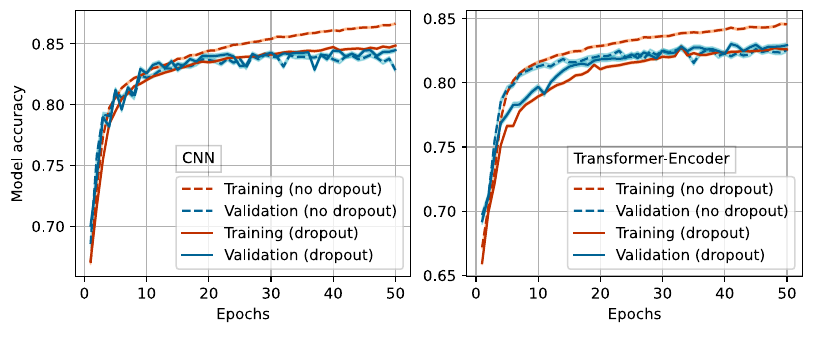}
    \caption{Learning curves for the Convolutional Neural Network (left) and the Transformer-Encoder (right), considering $w=1$ mm and $E_t=50$ keV. The dashed lines show the performance in the absence of dropout layers, while the solid lines include the effect of dropout layers (inserted accordingly to Fig. \ref{fig:3} and \ref{fig:4}), with a dropout rate of $0.15$ and $0.02$ for the CNN and the Transfomer-Encoder, respectively. We observe that in both cases the usage of dropout layers mitigates overfitting without compromising the asymptotical validation accuracy.}
    \label{fig:5}
\end{figure}

We note that in the case of no dropout layers, both models exhibit significant overfitting, approximatively after $15$-$20$ epochs, especially at small pixel size ($w$). Including dropout layers in our architecture is enough to minimise the occurrence of overfitting to negligible levels. We also observe that a dropout rate of $0.15$ is needed for the CNN, while for the Transformer-Encoder, overfitting is prevented with a rate as small as $0.02$. An example is shown in Fig. \ref{fig:5}.
\begin{figure*}
    \subfloat[\label{fig:6a}]{%
        \includegraphics[width=0.49\textwidth]{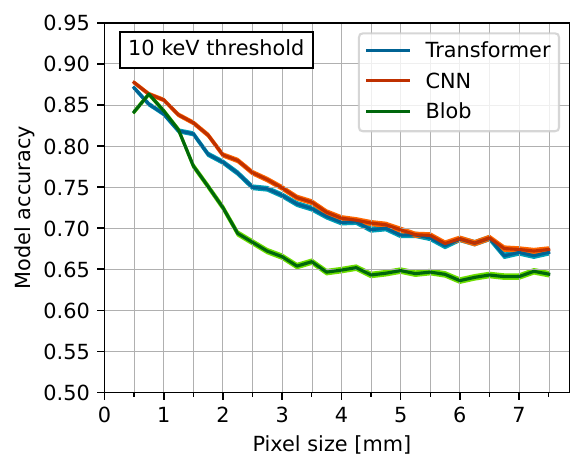}
    }
    \hfill
    \subfloat[\label{fig:6b}]{%
        \includegraphics[width=0.49\textwidth]{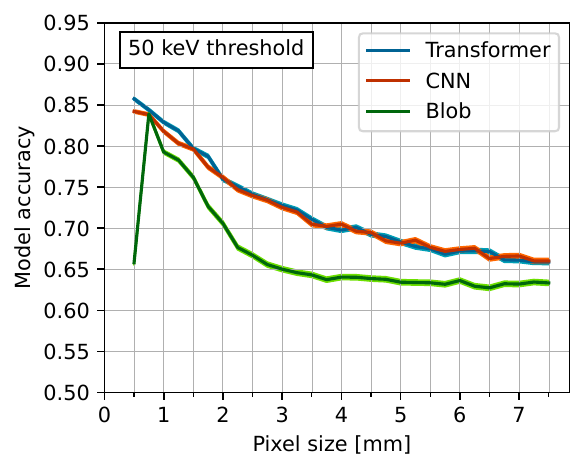}
    }
    \hfill
    \subfloat[\label{fig:6c}]{%
      \includegraphics[width=0.49\textwidth]{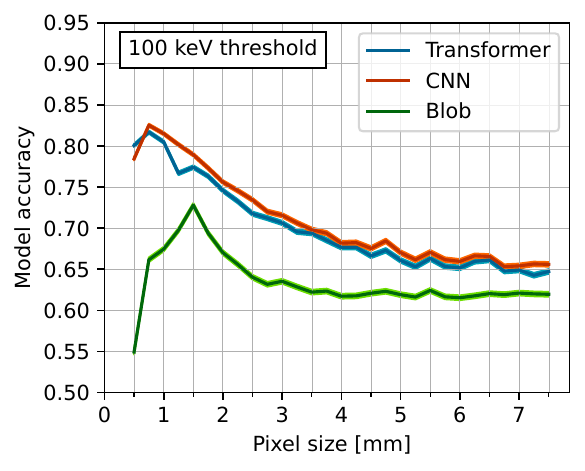}
    }
    \hfill
    \subfloat[\label{fig:6d}]{%
        \includegraphics[width=0.49\textwidth]{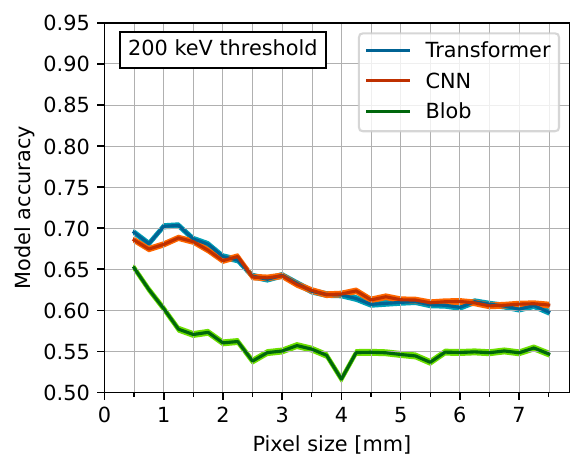}
}
\caption{Accuracy of the Transformer-Encoder, the Convolutional Neural Network and the blob method at different pixel sizes $w$ with a $0.25$ mm step applying a hit cutoff energy of $E_t=10$ keV (a), $50$ keV (b), $100$ keV (c), $200$ keV (d). The error bands account for $\pm 2\sigma$ statistical fluctuations.}
\label{fig:6}
\end{figure*}

Fig. \ref{fig:6} shows the classification accuracies of the blob method, the CNN, and the Transformer-Encoder for the test set at different pixel sizes and energy threshold values of $10$, $50$, $100$, and $200$ keV. It is important to notice that the average hit energy content scales with its volume. Since we keep one dimension fixed (as we only vary the pixel size and not the sampling frequency), the hit energies scale with $w^2$. As a consequence, the energy threshold has a bigger effect at low $w$. 

Overall, the blob model is unable to reach Deep Learning-competitive performances except for a handful of configurations with low-energy cutoffs and small pixel size ($w$ between $0.50$ mm and $1.25$ mm at $10$ keV, $w=0.75$ mm at $50$ keV). As expected, the blob model performs at its best when tracks are fine-grained and most of the hits pass the energy threshold. The presence of gaps in the LArTPC track reconstruction severely affects the graph connections described in Sec. \ref{sec:machine_learning} and compromises the blob candidate localization through the BFS algorithm. ML algorithms, instead, are much more robust to gaps in traces, showing little to no accuracy losses where the blob model fails.

Transformer-Encoder and CNN present a similar trend, especially at intermediate to high pixel sizes. Their behaviour differentiates at the $1$-$2\%$ level at $w\lesssim 2.5$ mm for $E_{t}=10$ keV and $E_{t}=100$ keV (CNN outperforms here) and $w\lesssim 1.5$ mm for $E_{t}=50$ keV and $E_{t}=200$ (Transformer-Encoder performs better). This trend is justified by the fact that more information is available when the number of hits increases and the effectiveness of the Transformer-Encoder learning method results in slightly better accuracy. We expect to gain further improvements in the classification accuracy for both models at small $w$, employing a larger training dataset and \textit{ad hoc} architecture and hyperparameter optimization for each individual $(w,\, E_t)$ configuration. For larger values of $w$, the analysis approaches the one-dimensional limit, as the time-axis alone carries most of the information, narrowing the improvement margin.

The results also emphasize how the accuracy dependence on $w$ flattens as $E_t$ increases. For $E_t=200$ keV, reducing $w$ from $w=7.5$ mm to $w=0.5$ mm -- a substantial increase of the LArTPC complexity and cost -- the rate of correctly classified events improves by just $10\%$, going from $60\%$ to $70\%$. 

\begin{figure*}
    \centering
    \subfloat[$w=1$ mm, $E_t=50$ keV.\label{fig:7a}]{%
        \includegraphics[width=0.48\textwidth]{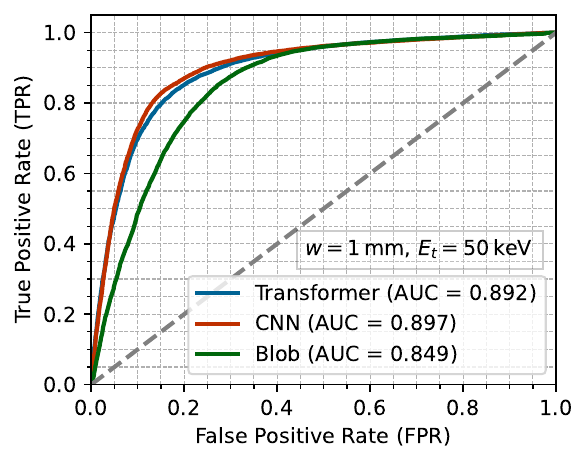}
    }
    \hfill
    \vspace{0.5cm}
    \subfloat[$w=5$ mm, $E_t=200$ keV. \label{fig:7b}]{%
        \includegraphics[width=0.48\textwidth]{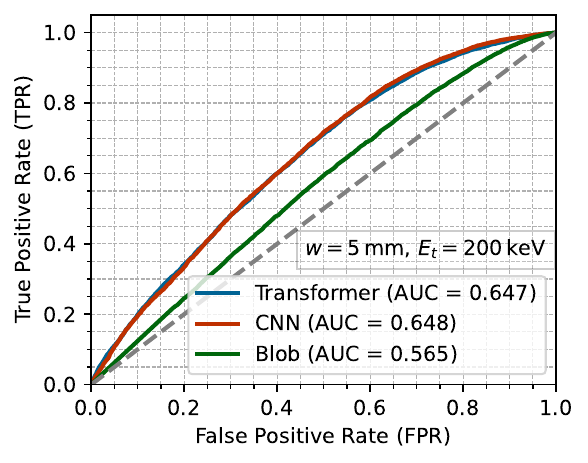}
    }
    \caption{ROC curves on a test dataset for Transformer-Encoder, CNN and blob models, displaying the tradeoff between the True Positive Ratio, defined as the fraction of correctly classified $\beta\beta$ events over the totality of them, and the False Positive Rate which corresponds to the fraction of $\beta$ events erroneously classified as $\beta\beta$. The two panels refer to different resolution-threshold conditions.}
    \label{fig:7}
\end{figure*}

Fig. \ref{fig:7} shows the Receiver Operating Characteristic (ROC) curves for each of the classification models and the corresponding Area Under Curve (AUC) evaluation metric \cite{BRADLEY19971145}, providing more complete information on two significant granularity-thresh\-old combinations ($w=5$ mm, $E_t=200$ keV and $w=1$ mm, $E_t=50$ keV) in terms of tradeoffs between signal and background acceptances. The first one (Fig.\ref{fig:7a}) corresponds to a regime comparable to the ones expected for FD1 HD and FD2 VD module, while the second one (Fig.\ref{fig:7b}) to an optimal, yet achievable configuration (see Sec.\ref{sec:physics}). In the first scenario, the ML techniques exhibit an overall superior performance with respect to the blob method. However, when setting a working point with high $\beta\beta$ selection efficiency (resulting, in turn, in a lower background rejection rate), all three models demonstrate comparable capabilities. Such a working point is the typical choice when searching for rare events in the presence of a dominant background, as would be the $0\nu\beta\beta$ process. Conversely, in the lower-resolution, high-threshold scenario, the ML models consistently outperform the blob method across all working points.

Machine learning techniques offer additional insight with respect to classic algorithms when detector optimization studies are considered.
Both the CNN and Transformer-based classification algorithms point toward the prominence of readout electronics over granularity for sufficiently small values of $w$. This is an important finding since the increase of granularity in large-volume LArTPC is a major technical challenge. Such an increase does not represent a viable option for $w<1$ mm due to the increase in the number of channels and the data throughput, while a reduction of $E_t$ at the level of a few tens of keV is well within reach of current technologies.

At $E_t=100$ keV and lower, the improvement in accuracy achievable by lowering $w$ is prominent and drives the accuracy metric of the $\beta$ vs $\beta \beta $ classification. 

\section{Conclusions}
\label{sec:conclusions}

In this paper, we discussed the performance of two major classes of machine learning algorithms for the identification of low-energy events in liquid argon and compared these findings with the performance of conventional techniques.  In particular, we focused our attention on the most challenging classification problem, the discrimination of single $\beta$ versus $\beta \beta$ events. We thus used as a benchmark the {\it blob} deterministic method developed by the NEXT Collaboration for the identification of neutrinoless double-beta decay modified to operate in a LArTPC. Both classes of machine learning algorithms - Convolutional Neural Networks and 
Transformer-Encoder - outperforms the blob algorithm. The CNN and transformer performance are comparable in most of the detector parameter space (see Fig. \ref{fig:6}). Overfitting is mitigated in both cases by a dropout layer and is negligible even for small values of the pixel size $w$. Still, the Transformer-Encoder is more memory-efficient and robust against overfitting even with a dropout rate as small as $0.02$.
ML-assisted techniques are particularly effective for detector optimization studies since the redefinition of conventional algorithms in a broad detector parameter phase space is very cumbersome. 
The CNN and Transformer-based classification algorithms point toward the prominence of readout electronics over granularity for sufficiently small values of $w$. This is an important finding since the increase of granularity in large-volume LArTPC is a major technical challenge, while a reduction of $E_t$ at the level of a few tens of keV is well within reach of current technologies.

\backmatter

\bmhead{Acknowledgments}

We gratefully acknowledge the CIEMAT neutrino group and, in particular, C. Cuesta, I. Gil Botella, A. Alvarez Garrote and P. Barham Alzás, for many useful discussions on the DUNE low-energy physics.

\section*{Declarations}

\begin{itemize}
\item Funding: MG and SV are supported by CERN through the CERN QTI. AG acknowledges support by the Horizon 2020 Marie Sk\l{}odowska-Curie actions (H2020-MSCA-IF GA No.101027746). This work is supported in part by the Italian MIUR Grant 2017-NAZ-0444 (PRIN2017). 
\item Conflict of interest/Competing interests: no conflicts to declare.
\item Availability of data and materials: the dataset generated and analysed during this study is available upon reasonable request from the author.
\item Code availability: the results presented in this manuscript have been produced with the DeepLAr software package available on the GitHub repository: \url{https://github.com/CERN-IT-INNOVATION/DeepLAr}
\end{itemize}

\noindent

\bibliography{bibliography}


\begin{thebibliography}{63}
\ifx \bisbn   \undefined \def \bisbn  #1{ISBN #1}\fi
\ifx \binits  \undefined \def \binits#1{#1}\fi
\ifx \bauthor  \undefined \def \bauthor#1{#1}\fi
\ifx \batitle  \undefined \def \batitle#1{#1}\fi
\ifx \bjtitle  \undefined \def \bjtitle#1{#1}\fi
\ifx \bvolume  \undefined \def \bvolume#1{\textbf{#1}}\fi
\ifx \byear  \undefined \def \byear#1{#1}\fi
\ifx \bissue  \undefined \def \bissue#1{#1}\fi
\ifx \bfpage  \undefined \def \bfpage#1{#1}\fi
\ifx \blpage  \undefined \def \blpage #1{#1}\fi
\ifx \burl  \undefined \def \burl#1{\textsf{#1}}\fi
\ifx \doiurl  \undefined \def \doiurl#1{\url{https://doi.org/#1}}\fi
\ifx \betal  \undefined \def \betal{\textit{et al.}}\fi
\ifx \binstitute  \undefined \def \binstitute#1{#1}\fi
\ifx \binstitutionaled  \undefined \def \binstitutionaled#1{#1}\fi
\ifx \bctitle  \undefined \def \bctitle#1{#1}\fi
\ifx \beditor  \undefined \def \beditor#1{#1}\fi
\ifx \bpublisher  \undefined \def \bpublisher#1{#1}\fi
\ifx \bbtitle  \undefined \def \bbtitle#1{#1}\fi
\ifx \bedition  \undefined \def \bedition#1{#1}\fi
\ifx \bseriesno  \undefined \def \bseriesno#1{#1}\fi
\ifx \blocation  \undefined \def \blocation#1{#1}\fi
\ifx \bsertitle  \undefined \def \bsertitle#1{#1}\fi
\ifx \bsnm \undefined \def \bsnm#1{#1}\fi
\ifx \bsuffix \undefined \def \bsuffix#1{#1}\fi
\ifx \bparticle \undefined \def \bparticle#1{#1}\fi
\ifx \barticle \undefined \def \barticle#1{#1}\fi
\bibcommenthead
\ifx \bconfdate \undefined \def \bconfdate #1{#1}\fi
\ifx \botherref \undefined \def \botherref #1{#1}\fi
\ifx \url \undefined \def \url#1{\textsf{#1}}\fi
\ifx \bchapter \undefined \def \bchapter#1{#1}\fi
\ifx \bbook \undefined \def \bbook#1{#1}\fi
\ifx \bcomment \undefined \def \bcomment#1{#1}\fi
\ifx \oauthor \undefined \def \oauthor#1{#1}\fi
\ifx \citeauthoryear \undefined \def \citeauthoryear#1{#1}\fi
\ifx \endbibitem  \undefined \def \endbibitem {}\fi
\ifx \bconflocation  \undefined \def \bconflocation#1{#1}\fi
\ifx \arxivurl  \undefined \def \arxivurl#1{\textsf{#1}}\fi
\csname PreBibitemsHook\endcsname

\bibitem[\protect\citeauthoryear{Abi et~al.}{2020a}]{DUNE:2020lwj}
\begin{barticle}
\bauthor{\bsnm{Abi}, \binits{B.}}, \betal:
\batitle{{Deep Underground Neutrino Experiment (DUNE), Far Detector Technical Design Report, Volume I Introduction to DUNE}}.
\bjtitle{JINST}
\bvolume{15}(\bissue{08}),
\bfpage{08008}
(\byear{2020})
\doiurl{10.1088/1748-0221/15/08/T08008}
{\href{https://arxiv.org/abs/2002.02967}{{arXiv:2002.02967}}}
{[physics.ins-det]}
\end{barticle}
\endbibitem

\bibitem[\protect\citeauthoryear{Abi et~al.}{2020b}]{DUNE:2020txw}
\begin{barticle}
\bauthor{\bsnm{Abi}, \binits{B.}}, \betal:
\batitle{{Deep Underground Neutrino Experiment (DUNE), Far Detector Technical Design Report, Volume IV: Far Detector Single-phase Technology}}.
\bjtitle{JINST}
\bvolume{15}(\bissue{08}),
\bfpage{08010}
(\byear{2020})
\doiurl{10.1088/1748-0221/15/08/T08010}
{\href{https://arxiv.org/abs/2002.03010}{{arXiv:2002.03010}}}
{[physics.ins-det]}
\end{barticle}
\endbibitem

\bibitem[\protect\citeauthoryear{Aalseth et~al.}{2018}]{DarkSide-20k:2017zyg}
\begin{barticle}
\bauthor{\bsnm{Aalseth}, \binits{C.E.}}, \betal:
\batitle{{DarkSide-20k: A 20 tonne two-phase LAr TPC for direct dark matter detection at LNGS}}.
\bjtitle{Eur. Phys. J. Plus}
\bvolume{133},
\bfpage{131}
(\byear{2018})
\doiurl{10.1140/epjp/i2018-11973-4}
{\href{https://arxiv.org/abs/1707.08145}{{arXiv:1707.08145}}}
{[physics.ins-det]}
\end{barticle}
\endbibitem

\bibitem[\protect\citeauthoryear{Agnes et~al.}{2015}]{DarkSide:2014llq}
\begin{barticle}
\bauthor{\bsnm{Agnes}, \binits{P.}}, \betal:
\batitle{{First Results from the DarkSide-50 Dark Matter Experiment at Laboratori Nazionali del Gran Sasso}}.
\bjtitle{Phys. Lett. B}
\bvolume{743},
\bfpage{456}--\blpage{466}
(\byear{2015})
\doiurl{10.1016/j.physletb.2015.03.012}
{\href{https://arxiv.org/abs/1410.0653}{{arXiv:1410.0653}}}
{[astro-ph.CO]}
\end{barticle}
\endbibitem

\bibitem[\protect\citeauthoryear{Adamowski et~al.}{2014}]{Adamowski:2014daa}
\begin{barticle}
\bauthor{\bsnm{Adamowski}, \binits{M.}}, \betal:
\batitle{{The Liquid Argon Purity Demonstrator}}.
\bjtitle{JINST}
\bvolume{9},
\bfpage{07005}
(\byear{2014})
\doiurl{10.1088/1748-0221/9/07/P07005}
{\href{https://arxiv.org/abs/1403.7236}{{arXiv:1403.7236}}}
{[physics.ins-det]}
\end{barticle}
\endbibitem

\bibitem[\protect\citeauthoryear{Montanari et~al.}{2015a}]{LBNE:2013lpy}
\begin{barticle}
\bauthor{\bsnm{Montanari}, \binits{D.}}, \betal:
\batitle{{First scientific application of the membrane cryostat technology.}}
\bjtitle{AIP Conf. Proc.}
\bvolume{1573}(\bissue{1}),
\bfpage{1664}--\blpage{1671}
(\byear{2015})
\doiurl{10.1063/1.4860907}
\end{barticle}
\endbibitem

\bibitem[\protect\citeauthoryear{Montanari et~al.}{2015b}]{Montanari:2015zwa}
\begin{barticle}
\bauthor{\bsnm{Montanari}, \binits{D.}},
\bauthor{\bsnm{Adamowski}, \binits{M.}},
\bauthor{\bsnm{Hahn}, \binits{A.}},
\bauthor{\bsnm{Norris}, \binits{B.}},
\bauthor{\bsnm{Reichenbacher}, \binits{J.}},
\bauthor{\bsnm{Rucinski}, \binits{R.}},
\bauthor{\bsnm{Stewart}, \binits{J.}},
\bauthor{\bsnm{Tope}, \binits{T.}}:
\batitle{{Performance and Results of the LBNE 35 Ton Membrane Cryostat Prototype}}.
\bjtitle{Phys. Procedia}
\bvolume{67},
\bfpage{308}--\blpage{313}
(\byear{2015})
\doiurl{10.1016/j.phpro.2015.06.092}
\end{barticle}
\endbibitem

\bibitem[\protect\citeauthoryear{Abi et~al.}{2020}]{DUNE:2020cqd}
\begin{barticle}
\bauthor{\bsnm{Abi}, \binits{B.}}, \betal:
\batitle{{First results on ProtoDUNE-SP liquid argon time projection chamber performance from a beam test at the CERN Neutrino Platform}}.
\bjtitle{JINST}
\bvolume{15}(\bissue{12}),
\bfpage{12004}
(\byear{2020})
\doiurl{10.1088/1748-0221/15/12/P12004}
{\href{https://arxiv.org/abs/2007.06722}{{arXiv:2007.06722}}}
{[physics.ins-det]}
\end{barticle}
\endbibitem

\bibitem[\protect\citeauthoryear{Agnes et~al.}{2021}]{DarkSide-20k:2021nia}
\begin{barticle}
\bauthor{\bsnm{Agnes}, \binits{P.}}, \betal:
\batitle{{Separating ${^{39} {Ar}}$ from ${^{40}{Ar}}$ by cryogenic distillation with Aria for dark-matter searches}}.
\bjtitle{Eur. Phys. J. C}
\bvolume{81}(\bissue{4}),
\bfpage{359}
(\byear{2021})
\doiurl{10.1140/epjc/s10052-021-09121-9}
{\href{https://arxiv.org/abs/2101.08686}{{arXiv:2101.08686}}}
{[physics.ins-det]}
\end{barticle}
\endbibitem

\bibitem[\protect\citeauthoryear{Alexander et~al.}{2019}]{Alexander:2019uvv}
\begin{botherref}
\oauthor{\bsnm{Alexander}, \binits{T.}}, et al.:
{The Low-Radioactivity Underground Argon Workshop: A workshop synopsis}
(2019).
\doiurl{10.48550/arXiv.1901.10108}
\end{botherref}
\endbibitem

\bibitem[\protect\citeauthoryear{Abed~Abud et~al.}{2022}]{DUNE:2022aul}
\begin{botherref}
\oauthor{\bsnm{Abed~Abud}, \binits{A.}}, et al.:
{Snowmass Neutrino Frontier: DUNE Physics Summary}
(2022).
\doiurl{10.48550/arXiv.2203.06100}
\end{botherref}
\endbibitem

\bibitem[\protect\citeauthoryear{Borkum et~al.}{2023}]{Borkum:2023dsu}
\begin{botherref}
\oauthor{\bsnm{Borkum}, \binits{A.}}, et al.:
{Large Low Background kTon-Scale Liquid Argon Time Projection Chambers}
(2023).
\doiurl{10.48550/arXiv.2301.11878}
\end{botherref}
\endbibitem

\bibitem[\protect\citeauthoryear{Back et~al.}{2022}]{Back:2022maq}
\begin{bchapter}
\bauthor{\bsnm{Back}, \binits{H.O.}}, \betal:
\bctitle{{A Facility for Low-Radioactivity Underground Argon}}.
In: \bbtitle{2022 Snowmass Summer Study}
(\byear{2022}).
\doiurl{10.48550/arXiv.2203.09734}
\end{bchapter}
\endbibitem

\bibitem[\protect\citeauthoryear{Avasthi et~al.}{2022}]{Avasthi:2022tjr}
\begin{bchapter}
\bauthor{\bsnm{Avasthi}, \binits{A.}}, \betal:
\bctitle{{Low Background kTon-Scale Liquid Argon Time Projection Chambers}}.
In: \bbtitle{2022 Snowmass Summer Study}
(\byear{2022}).
\doiurl{10.48550/arXiv.2203.08821}
\end{bchapter}
\endbibitem

\bibitem[\protect\citeauthoryear{Parsa et~al.}{2022}]{Parsa:2022mnj}
\begin{bchapter}
\bauthor{\bsnm{Parsa}, \binits{S.}}, \betal:
\bctitle{{SoLAr: Solar Neutrinos in Liquid Argon}}.
In: \bbtitle{2022 Snowmass Summer Study}
(\byear{2022}).
\doiurl{10.48550/arXiv.2203.07501}
\end{bchapter}
\endbibitem

\bibitem[\protect\citeauthoryear{Caratelli et~al.}{2022}]{caratelli2022lowenergy}
\begin{botherref}
\oauthor{\bsnm{Caratelli}, \binits{D.}},
\oauthor{\bsnm{Foreman}, \binits{W.}},
\oauthor{\bsnm{Friedland}, \binits{A.}},
\oauthor{\bsnm{Gardiner}, \binits{S.}},
\oauthor{\bsnm{Gil-Botella}, \binits{I.}},
\oauthor{\bsnm{al.}}:
Low-Energy Physics in Neutrino LArTPCs
(2022).
\doiurl{10.48550/arXiv.2203.00740}
\end{botherref}
\endbibitem

\bibitem[\protect\citeauthoryear{Abi et~al.}{2021}]{DUNE:2020zfm}
\begin{barticle}
\bauthor{\bsnm{Abi}, \binits{B.}}, \betal:
\batitle{{Supernova neutrino burst detection with the Deep Underground Neutrino Experiment}}.
\bjtitle{Eur. Phys. J. C}
\bvolume{81}(\bissue{5}),
\bfpage{423}
(\byear{2021})
\doiurl{10.1140/epjc/s10052-021-09166-w}
{\href{https://arxiv.org/abs/2008.06647}{{arXiv:2008.06647}}}
{[hep-ex]}
\end{barticle}
\endbibitem

\bibitem[\protect\citeauthoryear{Abi et~al.}{2020}]{DUNE:2020ypp}
\begin{botherref}
\oauthor{\bsnm{Abi}, \binits{B.}}, et al.:
{Deep Underground Neutrino Experiment (DUNE), Far Detector Technical Design Report, Volume II: DUNE Physics}
(2020).
\doiurl{10.48550/arXiv.2002.03005}
\end{botherref}
\endbibitem

\bibitem[\protect\citeauthoryear{Gil~Botella et~al.}{}]{moo_workshop_2022}
\begin{botherref}
\oauthor{\bsnm{Gil~Botella}, \binits{I.}}, et al.:
{DUNE Module of Opportunity Workshop}.
Valencia, 2-4 Nov 2022, \texttt{https://congresos.adeituv.es/dune\_science/}
\end{botherref}
\endbibitem

\bibitem[\protect\citeauthoryear{Capozzi et~al.}{2019}]{Capozzi:2018dat}
\begin{barticle}
\bauthor{\bsnm{Capozzi}, \binits{F.}},
\bauthor{\bsnm{Li}, \binits{S.W.}},
\bauthor{\bsnm{Zhu}, \binits{G.}},
\bauthor{\bsnm{Beacom}, \binits{J.F.}}:
\batitle{{DUNE as the Next-Generation Solar Neutrino Experiment}}.
\bjtitle{Phys. Rev. Lett.}
\bvolume{123}(\bissue{13}),
\bfpage{131803}
(\byear{2019})
\doiurl{10.1103/PhysRevLett.123.131803}
{\href{https://arxiv.org/abs/1808.08232}{{arXiv:1808.08232}}}
{[hep-ph]}
\end{barticle}
\endbibitem

\bibitem[\protect\citeauthoryear{Mastbaum et~al.}{2022}]{Mastbaum:2022rhw}
\begin{barticle}
\bauthor{\bsnm{Mastbaum}, \binits{A.}},
\bauthor{\bsnm{Psihas}, \binits{F.}},
\bauthor{\bsnm{Zennamo}, \binits{J.}}:
\batitle{{Xenon-doped liquid argon TPCs as a neutrinoless double beta decay platform}}.
\bjtitle{Phys. Rev. D}
\bvolume{106}(\bissue{9}),
\bfpage{092002}
(\byear{2022})
\doiurl{10.1103/PhysRevD.106.092002}
{\href{https://arxiv.org/abs/2203.14700}{{arXiv:2203.14700}}}
{[hep-ex]}
\end{barticle}
\endbibitem

\bibitem[\protect\citeauthoryear{Campestrini et~al.}{2014}]{CAMPESTRINI2014139}
\begin{barticle}
\bauthor{\bsnm{Campestrini}, \binits{M.}},
\bauthor{\bsnm{Stringari}, \binits{P.}},
\bauthor{\bsnm{Arpentinier}, \binits{P.}}:
\batitle{Solid–liquid equilibrium prediction for binary mixtures of ar, o2, n2, kr, xe, and ch4 using the lj-slv-eos}.
\bjtitle{Fluid Phase Equilibria}
\bvolume{379},
\bfpage{139}--\blpage{147}
(\byear{2014})
\end{barticle}
\endbibitem

\bibitem[\protect\citeauthoryear{Gallice}{2022}]{Gallice:2021ykz}
\begin{barticle}
\bauthor{\bsnm{Gallice}, \binits{N.}}:
\batitle{{Xenon doping of liquid argon in ProtoDUNE single phase}}.
\bjtitle{JINST}
\bvolume{17}(\bissue{01}),
\bfpage{01034}
(\byear{2022})
\doiurl{10.1088/1748-0221/17/01/C01034}
{\href{https://arxiv.org/abs/2111.00347}{{arXiv:2111.00347}}}
{[physics.ins-det]}
\end{barticle}
\endbibitem

\bibitem[\protect\citeauthoryear{Guffanti et~al.}{}]{ar42distillation}
\begin{botherref}
\oauthor{\bsnm{Guffanti}, \binits{D.}}, et al.:
{Depletion of atmospheric argon for neutrinoless double beta decay searches}.
In preparation
\end{botherref}
\endbibitem

\bibitem[\protect\citeauthoryear{Adams et~al.}{2020}]{Adams_2020}
\begin{barticle}
\bauthor{\bsnm{Adams}, \binits{C.}},
\bauthor{\bsnm{Tutto}, \binits{M.D.}},
\bauthor{\bsnm{Asaadi}, \binits{J.}},
\bauthor{\bsnm{Bernstein}, \binits{M.}},
\bauthor{\bsnm{al.}}:
\batitle{Enhancing neutrino event reconstruction with pixel-based 3d readout for liquid argon time projection chambers}.
\bjtitle{J. Instrum.}
\bvolume{15}(\bissue{04}),
\bfpage{04009}
(\byear{2020})
\doiurl{10.1088/1748-0221/15/04/P04009}
\end{barticle}
\endbibitem

\bibitem[\protect\citeauthoryear{Kubota et~al.}{2022}]{Q-Pix}
\begin{barticle}
\bauthor{\bsnm{Kubota}, \binits{S.}},
\bauthor{\bsnm{Ho}, \binits{J.}},
\bauthor{\bsnm{McDonald}, \binits{A.D.}},
\bauthor{\bsnm{Tata}, \binits{N.}},
\bauthor{\bsnm{Asaadi}, \binits{J.}},
\bauthor{\bsnm{al.}}:
\batitle{Enhanced low-energy supernova burst detection in large liquid argon time projection chambers enabled by q-pix}.
\bjtitle{Phys. Rev. D}
\bvolume{106},
\bfpage{032011}
(\byear{2022})
\doiurl{10.1103/PhysRevD.106.032011}
\end{barticle}
\endbibitem

\bibitem[\protect\citeauthoryear{Abi et~al.}{2020}]{DUNE:2020gpm}
\begin{barticle}
\bauthor{\bsnm{Abi}, \binits{B.}}, \betal:
\batitle{{Neutrino interaction classification with a convolutional neural network in the DUNE far detector}}.
\bjtitle{Phys. Rev. D}
\bvolume{102}(\bissue{9}),
\bfpage{092003}
(\byear{2020})
\doiurl{10.1103/PhysRevD.102.092003}
{\href{https://arxiv.org/abs/2006.15052}{{arXiv:2006.15052}}}
{[physics.ins-det]}
\end{barticle}
\endbibitem

\bibitem[\protect\citeauthoryear{Acciarri et~al.}{2021}]{SBNDGeV}
\begin{botherref}
\oauthor{\bsnm{Acciarri}, \binits{R.}},
\oauthor{\bsnm{Adams}, \binits{C.}},
\oauthor{\bsnm{Andreopoulos}, \binits{C.}},
\oauthor{\bsnm{Asaadi}, \binits{J.}},
\oauthor{\bsnm{al.}}:
Cosmic ray background removal with deep neural networks in sbnd.
Front. Artif. Intell.
\textbf{4}
(2021)
\doiurl{10.3389/frai.2021.649917}
\end{botherref}
\endbibitem

\bibitem[\protect\citeauthoryear{Abratenko et~al.}{2021}]{sparseCNN}
\begin{barticle}
\bauthor{\bsnm{Abratenko}, \binits{P.}},
\bauthor{\bsnm{Alrashed}, \binits{M.}},
\bauthor{\bsnm{An}, \binits{R.}},
\bauthor{\bsnm{Anthony}, \binits{J.}},
\bauthor{\bsnm{Asaadi}, \binits{J.}},
\bauthor{\bsnm{al.}}:
\batitle{Semantic segmentation with a sparse convolutional neural network for event reconstruction in microboone}.
\bjtitle{Phys. Rev. D}
\bvolume{103},
\bfpage{052012}
(\byear{2021})
\doiurl{10.1103/PhysRevD.103.052012}
\end{barticle}
\endbibitem

\bibitem[\protect\citeauthoryear{Adams et~al.}{2019}]{MicroBooNENN}
\begin{barticle}
\bauthor{\bsnm{Adams}, \binits{C.}},
\bauthor{\bsnm{Alrashed}, \binits{M.}},
\bauthor{\bsnm{An}, \binits{R.}},
\bauthor{\bsnm{Anthony}, \binits{J.}},
\bauthor{\bsnm{al.}}:
\batitle{Deep neural network for pixel-level electromagnetic particle identification in the microboone liquid argon time projection chamber}.
\bjtitle{Phys. Rev. D}
\bvolume{99},
\bfpage{092001}
(\byear{2019})
\doiurl{10.1103/PhysRevD.99.092001}
\end{barticle}
\endbibitem

\bibitem[\protect\citeauthoryear{Buuck et~al.}{2023}]{Buuck:2022duk}
\begin{barticle}
\bauthor{\bsnm{Buuck}, \binits{M.}},
\bauthor{\bsnm{Mishra}, \binits{A.}},
\bauthor{\bsnm{Charles}, \binits{E.}},
\bauthor{\bsnm{Di~Lalla}, \binits{N.}},
\bauthor{\bsnm{Hitchcock}, \binits{O.A.}},
\bauthor{\bsnm{Monzani}, \binits{M.E.}},
\bauthor{\bsnm{Omodei}, \binits{N.}},
\bauthor{\bsnm{Shutt}, \binits{T.}}:
\batitle{{Low-energy Electron-track Imaging for a Liquid Argon Time-projection-chamber Telescope Concept Using Probabilistic Deep Learning}}.
\bjtitle{Astrophys. J.}
\bvolume{942}(\bissue{2}),
\bfpage{77}
(\byear{2023})
\doiurl{10.3847/1538-4357/aca329}
{\href{https://arxiv.org/abs/2207.07805}{{arXiv:2207.07805}}}
{[astro-ph.IM]}
\end{barticle}
\endbibitem

\bibitem[\protect\citeauthoryear{Acciarri et~al.}{2019}]{ArgoNeuT}
\begin{barticle}
\bauthor{\bsnm{Acciarri}, \binits{R.}},
\bauthor{\bsnm{Adams}, \binits{C.}},
\bauthor{\bsnm{Asaadi}, \binits{J.}},
\bauthor{\bsnm{Baller}, \binits{B.}},
\bauthor{\bsnm{Bolton}, \binits{T.}},
\bauthor{\bsnm{al.}}:
\batitle{Demonstration of mev-scale physics in liquid argon time projection chambers using argoneut}.
\bjtitle{Phys. Rev. D}
\bvolume{99},
\bfpage{012002}
(\byear{2019})
\doiurl{10.1103/PhysRevD.99.012002}
\end{barticle}
\endbibitem

\bibitem[\protect\citeauthoryear{Albertsson et~al.}{2018}]{whitepaper_machinelearning}
\begin{botherref}
\oauthor{\bsnm{Albertsson}, \binits{K.}}, et al.:
Machine Learning in High Energy Physics Community White Paper.
arXiv
(2018).
\doiurl{10.48550/ARXIV.1807.02876}
\end{botherref}
\endbibitem

\bibitem[\protect\citeauthoryear{Antonello et~al.}{2014}]{Antonello:2014eha}
\begin{barticle}
\bauthor{\bsnm{Antonello}, \binits{M.}}, \betal:
\batitle{{Experimental observation of an extremely high electron lifetime with the ICARUS-T600 LAr-TPC}}.
\bjtitle{JINST}
\bvolume{9}(\bissue{12}),
\bfpage{12006}
(\byear{2014})
\doiurl{10.1088/1748-0221/9/12/P12006}
{\href{https://arxiv.org/abs/1409.5592}{{arXiv:1409.5592}}}
{[physics.ins-det]}
\end{barticle}
\endbibitem

\bibitem[\protect\citeauthoryear{Hewes et~al.}{2021}]{DUNE:2021tad}
\begin{barticle}
\bauthor{\bsnm{Hewes}, \binits{V.}}, \betal:
\batitle{{Deep Underground Neutrino Experiment (DUNE) Near Detector Conceptual Design Report}}.
\bjtitle{Instruments}
\bvolume{5}(\bissue{4}),
\bfpage{31}
(\byear{2021})
\doiurl{10.3390/instruments5040031}
{\href{https://arxiv.org/abs/2103.13910}{{arXiv:2103.13910}}}
{[physics.ins-det]}
\end{barticle}
\endbibitem

\bibitem[\protect\citeauthoryear{Adams et~al.}{2020}]{Adams:2020tqx}
\begin{barticle}
\bauthor{\bsnm{Adams}, \binits{D.}}, \betal:
\batitle{{The ProtoDUNE-SP LArTPC Electronics Production, Commissioning, and Performance}}.
\bjtitle{JINST}
\bvolume{15}(\bissue{06}),
\bfpage{06017}
(\byear{2020})
\doiurl{10.1088/1748-0221/15/06/P06017}
{\href{https://arxiv.org/abs/2002.01782}{{arXiv:2002.01782}}}
{[physics.ins-det]}
\end{barticle}
\endbibitem

\bibitem[\protect\citeauthoryear{Boulay and Hime}{2006}]{Boulay:2006mb}
\begin{barticle}
\bauthor{\bsnm{Boulay}, \binits{M.G.}},
\bauthor{\bsnm{Hime}, \binits{A.}}:
\batitle{{Technique for direct detection of weakly interacting massive particles using scintillation time discrimination in liquid argon}}.
\bjtitle{Astropart. Phys.}
\bvolume{25},
\bfpage{179}--\blpage{182}
(\byear{2006})
\doiurl{10.1016/j.astropartphys.2005.12.009}
\end{barticle}
\endbibitem

\bibitem[\protect\citeauthoryear{Andringa et~al.}{2023}]{Andringa:2023aax}
\begin{barticle}
\bauthor{\bsnm{Andringa}, \binits{S.}}, \betal:
\batitle{{Low-energy physics in neutrino LArTPCs}}.
\bjtitle{J. Phys. G}
\bvolume{50}(\bissue{3}),
\bfpage{033001}
(\byear{2023})
\doiurl{10.1088/1361-6471/acad17}
\end{barticle}
\endbibitem

\bibitem[\protect\citeauthoryear{Benetti et~al.}{2007}]{BENETTI200783}
\begin{barticle}
\bauthor{\bsnm{Benetti}, \binits{P.}},
\bauthor{\bsnm{Calaprice}, \binits{F.}},
\bauthor{\bsnm{Calligarich}, \binits{E.}},
\bauthor{\bsnm{Cambiaghi}, \binits{M.}},
\bauthor{\bsnm{Carbonara}, \binits{F.}},
\bauthor{\bsnm{al.}}:
\batitle{Measurement of the specific activity of 39ar in natural argon}.
\bjtitle{NIM-A}
\bvolume{574}(\bissue{1}),
\bfpage{83}--\blpage{88}
(\byear{2007})
\doiurl{10.1016/j.nima.2007.01.106}
\end{barticle}
\endbibitem

\bibitem[\protect\citeauthoryear{Ponkratenko et~al.}{2000}]{deacy4}
\begin{barticle}
\bauthor{\bsnm{Ponkratenko}, \binits{O.A.}},
\bauthor{\bsnm{Tretyak}, \binits{V.I.}},
\bauthor{\bsnm{Zdesenko}, \binits{Y.G.}}:
\batitle{{Event generator {DECAY}4 for simulating double-beta processes and decays of radioactive nuclei}}.
\bjtitle{Physics of Atomic Nuclei}
\bvolume{63}(\bissue{7}),
\bfpage{1282}--\blpage{1287}
(\byear{2000})
\doiurl{10.1134/1.855784}
\end{barticle}
\endbibitem

\bibitem[\protect\citeauthoryear{Agostinelli et~al.}{2003}]{Geant4:2003}
\begin{barticle}
\bauthor{\bsnm{Agostinelli}, \binits{S.}}, \betal:
\batitle{Geant4—a simulation toolkit}.
\bjtitle{Nucl. Instrum. Methods Phys. Res. A}
\bvolume{506}(\bissue{3}),
\bfpage{250}--\blpage{303}
(\byear{2003})
\doiurl{10.1016/S0168-9002(03)01368-8}
\end{barticle}
\endbibitem

\bibitem[\protect\citeauthoryear{Allison et~al.}{2006}]{Geant4:2006}
\begin{barticle}
\bauthor{\bsnm{Allison}, \binits{J.}}, \betal:
\batitle{Geant4 developments and applications}.
\bjtitle{IEEE Trans. Nucl. Sci.}
\bvolume{53}(\bissue{1}),
\bfpage{270}--\blpage{278}
(\byear{2006})
\doiurl{10.1109/TNS.2006.869826}
\end{barticle}
\endbibitem

\bibitem[\protect\citeauthoryear{Allison et~al.}{2016}]{Geant4:2016}
\begin{barticle}
\bauthor{\bsnm{Allison}, \binits{J.}}, \betal:
\batitle{Recent developments in geant4}.
\bjtitle{Nucl. Instrum. Methods Phys. Res. A}
\bvolume{835},
\bfpage{186}--\blpage{225}
(\byear{2016})
\doiurl{10.1016/j.nima.2016.06.125}
\end{barticle}
\endbibitem

\bibitem[\protect\citeauthoryear{Gomez-Cadenas}{2016}]{NEXT}
\begin{barticle}
\bauthor{\bsnm{Gomez-Cadenas}, \binits{J.J.}}:
\batitle{{The NEXT experiment}}.
\bjtitle{Nuclear and Particle Physics Proceedings}
\bvolume{273-275},
\bfpage{1732}--\blpage{1739}
(\byear{2016})
\doiurl{10.1016/j.nuclphysbps.2015.09.279}
\end{barticle}
\endbibitem

\bibitem[\protect\citeauthoryear{Mart{\'{\i}}n-Albo et~al.}{2016}]{Mart_n_Albo_2016}
\begin{botherref}
\oauthor{\bsnm{Mart{\'{\i}}n-Albo}, \binits{J.}},
\oauthor{\bsnm{Vidal}, \binits{J.M.}},
\oauthor{\bsnm{Ferrario}, \binits{P.}},
\oauthor{\bsnm{Nebot-Guinot}, \binits{M.}},
\oauthor{\bsnm{G{\'{o}}mez-Cadenas}, \binits{J.J.}}, et al.:
Sensitivity of {NEXT}-100 to neutrinoless double beta decay.
JHEP
\textbf{2016}(5)
(2016)
\doiurl{10.1007/jhep05(2016)159}
\end{botherref}
\endbibitem

\bibitem[\protect\citeauthoryear{Renner et~al.}{2017}]{JRenner}
\begin{botherref}
\oauthor{\bsnm{Renner}, \binits{J.}}, et al.:
Background rejection in next using deep neural networks.
JINST
\textbf{12}
(2017)
\doiurl{10.1088/1748-0221/12/01/T01004}
\end{botherref}
\endbibitem

\bibitem[\protect\citeauthoryear{O'Shea and Nash}{2015}]{CNN}
\begin{botherref}
\oauthor{\bsnm{O'Shea}, \binits{K.}},
\oauthor{\bsnm{Nash}, \binits{R.}}:
An Introduction to Convolutional Neural Networks.
arXiv
(2015).
\doiurl{10.48550/arXiv.1511.08458}
\end{botherref}
\endbibitem

\bibitem[\protect\citeauthoryear{Vaswani et~al.}{2017}]{aiayn}
\begin{botherref}
\oauthor{\bsnm{Vaswani}, \binits{A.}},
\oauthor{\bsnm{Shazeer}, \binits{N.}},
\oauthor{\bsnm{Parmar}, \binits{N.}},
\oauthor{\bsnm{Uszkoreit}, \binits{J.}},
\oauthor{\bsnm{Jones}, \binits{L.}},
\oauthor{\bsnm{Gomez}, \binits{A.N.}},
\oauthor{\bsnm{Kaiser}, \binits{L.}},
\oauthor{\bsnm{Polosukhin}, \binits{I.}}:
Attention Is All You Need.
arXiv
(2017).
\doiurl{10.48550/arXiv.1706.03762}
\end{botherref}
\endbibitem

\bibitem[\protect\citeauthoryear{Kekic et~al.}{2021}]{NEXT2021}
\begin{botherref}
\oauthor{\bsnm{Kekic}, \binits{M.}},
\oauthor{\bsnm{Adams}, \binits{C.}},
\oauthor{\bsnm{al.}}:
Demonstration of background rejection using deep convolutional neural networks in the next experiment.
J. High Energ. Phys.
\textbf{189}
(2021)
\doiurl{10.1007/JHEP01(2021)189}
\end{botherref}
\endbibitem

\bibitem[\protect\citeauthoryear{Lin et~al.}{2022}]{atnsurvey}
\begin{barticle}
\bauthor{\bsnm{Lin}, \binits{T.}},
\bauthor{\bsnm{Wang}, \binits{Y.}},
\bauthor{\bsnm{Liu}, \binits{X.}},
\bauthor{\bsnm{Qiu}, \binits{X.}}:
\batitle{A survey of transformers}.
\bjtitle{AI Open}
\bvolume{3},
\bfpage{111}--\blpage{132}
(\byear{2022})
\doiurl{10.1016/j.aiopen.2022.10.001}
\end{barticle}
\endbibitem

\bibitem[\protect\citeauthoryear{Workman and others [Particle Data~Group]}{2022}]{pdg}
\begin{barticle}
\bauthor{\bsnm{Workman}, \binits{R.L.}},
\bauthor{\bsnm{[Particle Data~Group]}}:
\batitle{{Review of Particle Physics}}.
\bjtitle{PTEP}
\bvolume{2022},
\bfpage{083}--\blpage{01}
(\byear{2022})
\doiurl{10.1093/ptep/ptac097}
\end{barticle}
\endbibitem

\bibitem[\protect\citeauthoryear{Silvela and Portillo}{2001}]{Silvela2001BreadthfirstSA}
\begin{barticle}
\bauthor{\bsnm{Silvela}, \binits{J.}},
\bauthor{\bsnm{Portillo}, \binits{J.}}:
\batitle{{Breadth-first search and its application to image processing problems}}.
\bjtitle{IEEE transactions on image processing : a publication of the IEEE Signal Processing Society}
\bvolume{10 8},
\bfpage{1194}--\blpage{9}
(\byear{2001})
\end{barticle}
\endbibitem

\bibitem[\protect\citeauthoryear{Neyman and Pearson}{1933}]{npl}
\begin{barticle}
\bauthor{\bsnm{Neyman}, \binits{J.}},
\bauthor{\bsnm{Pearson}, \binits{E.S.}}:
\batitle{{On the Problem of the Most Efficient Tests of Statistical Hypotheses}}.
\bjtitle{Philosophical Transactions of the Royal Society of London. Series A, Containing Papers of a Mathematical or Physical Character}
\bvolume{231},
\bfpage{289}--\blpage{337}
(\byear{1933})
\end{barticle}
\endbibitem

\bibitem[\protect\citeauthoryear{Bilal et~al.}{2018}]{Bilal_2018}
\begin{barticle}
\bauthor{\bsnm{Bilal}, \binits{A.}},
\bauthor{\bsnm{Jourabloo}, \binits{A.}},
\bauthor{\bsnm{Ye}, \binits{M.}},
\bauthor{\bsnm{Liu}, \binits{X.}},
\bauthor{\bsnm{Ren}, \binits{L.}}:
\batitle{Do convolutional neural networks learn class hierarchy?}
\bjtitle{{IEEE} Transactions on Visualization and Computer Graphics}
\bvolume{24}(\bissue{1}),
\bfpage{152}--\blpage{162}
(\byear{2018})
\doiurl{10.1109/tvcg.2017.2744683}
\end{barticle}
\endbibitem

\bibitem[\protect\citeauthoryear{Ioffe and Szegedy}{2015}]{bnorm}
\begin{botherref}
\oauthor{\bsnm{Ioffe}, \binits{S.}},
\oauthor{\bsnm{Szegedy}, \binits{C.}}:
Batch Normalization: Accelerating Deep Network Training by Reducing Internal Covariate Shift.
arXiv
(2015).
\doiurl{10.48550/arXiv.1502.03167}
\end{botherref}
\endbibitem

\bibitem[\protect\citeauthoryear{Srivastava et~al.}{2014}]{dropout}
\begin{barticle}
\bauthor{\bsnm{Srivastava}, \binits{N.}},
\bauthor{\bsnm{Hinton}, \binits{G.}},
\bauthor{\bsnm{Krizhevsky}, \binits{A.}},
\bauthor{\bsnm{Sutskever}, \binits{I.}},
\bauthor{\bsnm{Salakhutdinov}, \binits{R.}}:
\batitle{Dropout: A simple way to prevent neural networks from overfitting}.
\bjtitle{Journal of Machine Learning Research}
\bvolume{15}(\bissue{56}),
\bfpage{1929}--\blpage{1958}
(\byear{2014})
\end{barticle}
\endbibitem

\bibitem[\protect\citeauthoryear{Xu et~al.}{2015}]{leakyrelu}
\begin{botherref}
\oauthor{\bsnm{Xu}, \binits{B.}},
\oauthor{\bsnm{Wang}, \binits{N.}},
\oauthor{\bsnm{Chen}, \binits{T.}},
\oauthor{\bsnm{Li}, \binits{M.}}:
Empirical Evaluation of Rectified Activations in Convolutional Network.
arXiv
(2015).
\doiurl{10.48550/arXiv.1505.00853}
\end{botherref}
\endbibitem

\bibitem[\protect\citeauthoryear{Kouretas and Paliouras}{2020}]{softmax}
\begin{barticle}
\bauthor{\bsnm{Kouretas}, \binits{I.}},
\bauthor{\bsnm{Paliouras}, \binits{V.}}:
\batitle{{Hardware Implementation of a Softmax-Like Function for Deep Learning}}.
\bjtitle{Technologies}
\bvolume{8}(\bissue{3}),
\bfpage{46}
(\byear{2020})
\doiurl{10.3390/technologies8030046}
\end{barticle}
\endbibitem

\bibitem[\protect\citeauthoryear{Raffel and Ellis}{2016}]{contextvector1}
\begin{botherref}
\oauthor{\bsnm{Raffel}, \binits{C.}},
\oauthor{\bsnm{Ellis}, \binits{D.P.W.}}:
Feed-Forward Networks with Attention Can Solve Some Long-Term Memory Problems
(2016).
\doiurl{10.48550/arXiv.1512.08756}
\end{botherref}
\endbibitem

\bibitem[\protect\citeauthoryear{Brauwers and Frasincar}{2023}]{contextvector2}
\begin{barticle}
\bauthor{\bsnm{Brauwers}, \binits{G.}},
\bauthor{\bsnm{Frasincar}, \binits{F.}}:
\batitle{A general survey on attention mechanisms in deep learning}.
\bjtitle{IEEE Trans. Knowl. Data. Eng.}
\bvolume{35}(\bissue{4}),
\bfpage{3279}--\blpage{3298}
(\byear{2023})
\doiurl{10.1109/TKDE.2021.3126456}
\end{barticle}
\endbibitem

\bibitem[\protect\citeauthoryear{Li et~al.}{2016}]{Li_2016}
\begin{barticle}
\bauthor{\bsnm{Li}, \binits{Y.}},
\bauthor{\bsnm{Tsang}, \binits{T.}},
\bauthor{\bsnm{Thorn}, \binits{C.}},
\bauthor{\bsnm{Qian}, \binits{X.}},
\bauthor{\bsnm{Diwan}, \binits{M.}},
\bauthor{\bsnm{Joshi}, \binits{J.}},
\bauthor{\bsnm{Kettell}, \binits{S.}},
\bauthor{\bsnm{al.}}:
\batitle{Measurement of longitudinal electron diffusion in liquid argon}.
\bjtitle{Nucl. Instrum. Methods Phys. Res. A}
\bvolume{816},
\bfpage{160}--\blpage{170}
(\byear{2016})
\doiurl{10.1016/j.nima.2016.01.094}
\end{barticle}
\endbibitem

\bibitem[\protect\citeauthoryear{Kingma and Ba}{2017}]{kingma2017adam}
\begin{botherref}
\oauthor{\bsnm{Kingma}, \binits{D.P.}},
\oauthor{\bsnm{Ba}, \binits{J.}}:
Adam: A Method for Stochastic Optimization
(2017).
\doiurl{10.48550/arXiv.1412.6980}
\end{botherref}
\endbibitem

\bibitem[\protect\citeauthoryear{Bradley}{1997}]{BRADLEY19971145}
\begin{barticle}
\bauthor{\bsnm{Bradley}, \binits{A.P.}}:
\batitle{The use of the area under the roc curve in the evaluation of machine learning algorithms}.
\bjtitle{Pattern Recognit}
\bvolume{30}(\bissue{7}),
\bfpage{1145}--\blpage{1159}
(\byear{1997})
\doiurl{10.1016/S0031-3203(96)00142-2}
\end{barticle}
\endbibitem

\end{thebibliography}

\end{document}